\documentstyle[aps,epsfig]{revtex}

\def\beq{\begin{equation}}
\def\eeq{\end{equation}}
\def\beqn{\begin{eqnarray}}
\def\eeqn{\end{eqnarray}}

\begin{document}

\title{The GDH Sum Rule and Related Integrals}

\author{D. Drechsel, S.S. Kamalov\thanks{ Permanent address:
Laboratory of Theoretical Physics, JINR Dubna,
Head Post Office Box 79, SU-101000 Moscow, Russia.},
and L. Tiator}

\address{Institut f\"ur Kernphysik, Johannes Gutenberg-Universit\"at, D-55099
Mainz, Germany}

\maketitle

\begin{abstract}
The spin structure of the nucleon resonance region is analyzed on
the basis of our phenomenological model MAID. Predictions are
given for the Gerasimov-Drell-Hearn sum rule as well as
generalized integrals over spin structure functions. The
dependence of these integrals on momentum transfer is studied and
rigorous relationships between various definitions of generalized
Gerasimov-Drell-Hearn integrals and spin polarizabilities are
derived. These results are compared to the predictions of chiral
perturbation theory and phenomenological models.

\end{abstract}

\section{introduction}

Since the end of the 1970's, the spin structure of the nucleon has
been studied by scattering polarized lepton beams off polarized
targets. The aim of such experiments has been to measure the spin
structure functions $ g_{1}(x,Q^2)$ and $ g_{2}(x,Q^2)$, depending
on the fractional momentum of the constituents, $ x=Q^2/2m\nu$,
and the square of four-momentum transfer, $ q^2 = -Q^2 $, where
$\nu$ is the (virtual) photon lab energy and $m$ the nucleon mass.
Already the first experiments at CERN~\cite{Ash89} and
SLAC~\cite{Bau83} sparked considerable interest in the community,
because the first moment of $g_{1}$ , $\Gamma_{1}(Q^2) =
\int\limits_{0}^{1} g_{1}(x,Q^2) dx$, turned out to be
substantially below the quark model prediction~\cite{Ell74}. This
"spin crisis" led to the conclusion that less than half of the
nucleon spin is carried by the quarks. However, the difference of
the proton and neutron moments, $\Gamma_{1}^{p} - \Gamma_{1}^{n}$
, was found to be well described by Bjorken's sum
rule~\cite{Bjo66}, which is a strict QCD prediction.

In spite of new and more sophisticated experiments~\cite{Ada97},
the question about the carriers of the spin is still open. Various
new experimental proposals are presently underway at HERMES,
COMPASS and Jlab~\cite{Lam00}, and in the context of the ELFE
project~\cite{Van00a}. The idea behind these experiments is to
measure the so-called skewed parton distributions, which are
expected to reveal information on details of the parton structure,
such as sea quark, gluon and orbital momentum contributions to the
spin. The experimental tool to determine these new structure
functions will be semi-inclusive reactions such as deeply virtual
Compton scattering and the production of various mesons as filters
of particular quantum numbers~\cite{Van00a}.

Recent improvements in polarized beam and target techniques have
made it possible to determine the spin structure functions over an
increased range of kinematical values. In particular the E143
Collaboration at SLAC~\cite{Abe98} and the HERMES Collaboration at
DESY~\cite{Ack98} have obtained data at momentum transfers down to
$Q^2 \simeq 1$ (GeV/c)$^2 $. The range of momentum transfer below
these values will be covered by various experiments at the
Jefferson Lab, which are being performed in the full range $ 0.02
(GeV/c)^2 \leq Q^2 \leq 2$ (GeV/c)$^2 $~\cite{Bur91}. The first
direct experimental data for real photons $(Q^2 = 0)$ were
recently taken at MAMI~\cite{Ahr00} in the energy range 200~MeV$<
\nu < 800$~MeV, and data at the higher energies are expected from
ELSA within short.

In conclusion we expect that the spin structure functions will
soon be known over the full kinematical range. This will make it
possible to study the transition from the non-perturbative region
at low $Q^2$ to the perturbative region at large $Q^2$. In
particular the first moment $\Gamma_{1}$ is constrained, in the
limit of $Q^2 \rightarrow 0$ (real photons), by the famous
Gerasimov-Drell-Hearn sum rule (GDH, Ref.~\cite{Ger65}),
$\Gamma_{1} \rightarrow -Q^2\kappa^2_N/8m^2$ , where $\kappa_N$ is
the anomalous magnetic moment of the nucleon. The reader should
note that here and in the following we have included the inelastic
contributions to $\Gamma_{1}$ only. As has been pointed out by Ji
and Melnitchouk~\cite{JiM97}, the elastic contribution is in fact
the dominant one at small $Q^2$ and has to be taken into account
in comparing with twist expansions about the deep inelastic limit.

The GDH sum rule predicts $\Gamma_{1} < 0$ for small $Q^2$, while
all experiments for $Q^2 > 1$~(GeV/c)$^2$ yield clearly positive
values for the proton. Therefore, the spin structure has to change
rapidly at low $Q^2$, with some zero-crossing at $Q^2 <
1$~(GeV/c)$^2$. The expected strong variation of $\Gamma_1$ with
momentum transfer marks the transition from the physics of
resonance-driven coherent processes to incoherent scattering off
the constituents. The evolution of the sum rule was first
described by Anselmino et al.~\cite{Ans89} in terms of a
parametrization based on vector meson dominance. Burkert, Ioffe
and others~\cite{Bur92,Azn95,Sch99} refined this model considerably by
treating the contributions of the resonances explicitly. Soffer
and Teryaev~\cite{Sof93} suggested that the rapid fluctuation of
$\Gamma_{1}$ should be analyzed in conjunction with $\Gamma_{2}$ ,
the first moment of the second spin structure function. The latter
is constrained by the less familiar Burkhardt-Cottingham sum rule
(BC, Ref.~\cite{Bur70}) at all values of $Q^2$. Therefore the sum
of the two moments, $\Gamma_3=\Gamma_{1} + \Gamma_{2}$, is known
for both $Q^2 = 0$ and $Q^2 \rightarrow \infty$. Though this sum
is related to the practically unknown longitudinal-transverse
interference cross section $\sigma'_{LT}$ and therefore not yet
determined directly, Ref.~\cite{Sof93} assumed that it can be
extrapolated smoothly between the two limiting values of $Q^2$.
The rapid fluctuation of $\Gamma_{1}$ then follows by subtraction
of the BC value from $\Gamma_{3}$.

The small momentum evolution of a generalized GDH integral was
investigated by Bernard et al.~\cite{Ber93} in the framework of
heavy baryon ChPT. At $O(p^3)$ these authors predicted a positive
slope of this integral at $Q^2 = 0$, while the phenomenological
analysis of Burkert et al.~\cite{Bur92} indicated a negative
slope. These generalized GDH integrals contain information on both
spin structure functions, which are combined such that the
practically unknown longitudinal-transverse interference term
cancels. As will be explained later in detail, the definition of
these integrals in the literature is not unique.

Recently, Ji et al.~\cite{JiO99} have extended the calculations to
$O(p^{4})$. They find strong modifications due to the next-order term,
which even change the sign of the slope at the origin to negative
values much below the phenomenological analysis. In a similar way the
related forward spin polarizability $\gamma_{0}$ is changed
substantially by going from $O(p^3)$ to $O(p^4)$ , which may cast some
doubt on the convergence of the perturbation
series~\cite{Ber95,JiK99,Kum99,Gel00}. Unfortunately there appears an
additional problem concerning the decomposition of the Compton
amplitude in the nucleon pole terms (contained in the real Born
amplitude) and contributions of intermediate excited states (the
complex residual amplitude). The origin of the problem is due to the
Foldy-Wouthuysen type expansion in HBChPT, which changes the pole
structure at any given expansion in the nucleon mass $m$.

It will be the aim of this contribution to present our predictions for
the helicity-dependent cross sections, and to compare our results with
the existing data and other theoretical predictions. For this purpose
we shall review the formalism in sect.~2, with special emphasis on sum
rules and generalized GDH integrals. Our predictions will be compared
to the data and previous calculations in sect.~3, and we shall close
by a brief summary of our results in sect.~4.

\section{Formalism}

We consider the scattering of polarized electrons off polarized
target nucleons. The $lab$ energies of the electrons in the
initial and final states are denoted with $E$ and $E'$,
respectively. The incoming electrons carry the (longitudinal)
polarization $h=\pm 1$, and the two relevant polarization
components of the target are $P_z$ (parallel to the $lab$ momentum
$\vec{k}$ of the virtual photon) and $P_x$ (perpendicular to
$\vec{k}$ in the scattering plane of the electron and in the
half-plane of the outgoing electron). The differential cross
section for exclusive electroproduction can then be expressed in
terms of four ``virtual photoabsorption cross sections''
$\sigma_i(\nu, Q^2)$ by~\cite{Dre95,Dre99b}

\beq
\frac{d\sigma}{d\Omega\ dE'} = \Gamma \sigma (\nu,Q^2)\,,
\label{eq1}
\eeq

\beq
\sigma = \sigma_T+\epsilon\sigma_L+hP_x\sqrt{2\epsilon(1-\epsilon)}\
         \sigma'_{LT}+hP_z\sqrt{1-\epsilon^2}\sigma'_{TT}\, ,
\label{eq2}
\eeq
with

\beq
\Gamma = \frac{\alpha}{2\pi^2}\ \frac{E'}{E}\ \frac{K}{Q^2}\
         \frac{1}{1-\epsilon}
\label{eq3}
\eeq
the flux of the virtual photon field and $\epsilon$ its transverse
polarization, $\nu=E-E'$ the virtual photon energy in the $lab$
frame and $Q^2>0$ describing the square of the virtual photon
four-momentum. In accordance with our previous
notation~\cite{Dre99a} we shall define the flux with the ``photon
equivalent energy'' $K=K_H=(W^2-m^2)/2m=\nu(1-x)$, where $W$ is
the total $cm$ energy, $m$ the mass of the target nucleon, and
$x=Q^2/2m\nu$ the Bjorken scaling variable. We note that our
choice of the flux factor is originally due to Hand~\cite{Han63}.
Another often used definition was given by Gilman~\cite{Gil68} who
used $K=K_G=\nu\sqrt{1+\gamma^2}$, the $lab$ momentum of the
virtual photon, with $\gamma=Q/\nu$.

The quantities $\sigma_T$ and $\sigma'_{TT}$ can be expressed in
terms of the total cross sections $\sigma_{3/2}$ and
$\sigma_{1/2}$, corresponding to excitation of intermediate states
with spin projections $3/2$ and $1/2$, respectively,

\beq
\sigma_T = \frac{1}{2}(\sigma_{3/2} + \sigma_{1/2})\,,\qquad
\sigma'_{TT} = \frac{1}{2} (\sigma_{3/2}-\sigma_{1/2})\ .
\label{eq4}
\eeq

The virtual photoabsorption cross sections in Eq.~(\ref{eq2}) are
related to the quark structure functions $F_1$, $F_2$, $g_1$, and
$g_2$ depending on $\nu$ and $Q^2$,

\begin{eqnarray}
\label{eq5} \sigma_T& =& \frac{4\pi^2\alpha}{mK}\ F_1\ ,\nonumber
\\ \sigma_L& =& \frac{4\pi^2\alpha}{K}\left[ \frac{F_2}{\nu}
(1+\gamma^2) - \frac{F_1}{m}\right]\ ,\nonumber \\ \sigma'_{LT}&
=& -\frac{4\pi^2\alpha}{mK}\,\gamma\,(g_1+g_2)\ ,\nonumber \\
\sigma'_{TT}& =& -\frac{4\pi^2\alpha}{mK}\left (g_1-\gamma^2\,
g_2\right )\ .
\end{eqnarray}
In comparing with the standard nomenclature of deep inelastic
scattering DIS~\cite{Abe98} we note that
$\sigma'_{LT}=-\sigma_{LT}(DIS)$ and
$\sigma'_{TT}=-\sigma_{TT}(DIS)$. It is obvious that the virtual
absorption cross sections $\sigma_i$ depend on the choice of the
flux factor $K$. In the following we shall use the definition of
Hand, $K=K_H$. If we compare with the work of authors using the
convention of Gilman, $K=K_G$, our photoabsorption cross sections
of Eqs.~(\ref{eq5}) have to be multiplied by the ratio
$K_H/K_G=(1-x)/\sqrt{1+\gamma^2}$, i.e.
$\sigma_i^G=(1-x)(1+\gamma^2)^{-1/2}\sigma_i^H$.

We generalize the Gerasimov-Drell-Hearn (GDH) sum rule~\cite{Ger65} by
introducing the $Q^2$-dependent integral \beq I_1(Q^2) =
\frac{2m^2}{Q^2}\int_{0}^{x_0}g_1(x,Q^2)\ dx \rightarrow \left \{
  \begin{array}{lll} - \frac{1}{4}\kappa_N^2 &{\rm for}
    &Q^2\rightarrow 0\\ \frac{2m^2}{Q^2}\Gamma_1 + {\cal O}(Q^{-4})
    &{\rm for} &Q^2\rightarrow \infty
\end{array} \right. \,,
\label{eq6}
\eeq
where $x_0= Q^2/(2mm_{\pi}+m_{\pi}^2+Q^2)$ is the threshold for
one-pion production. In the scaling regime the structure functions
should depend on $x$ only, and $\Gamma_1=\int g_1(x) dx = const$.

For the second spin structure function the Burkhardt-Cottingham (BC)
sum rule asserts that the integral over $g_2$ vanishes if integrated
over both elastic and inelastic contributions~\cite{Bur70}. As a
consequence one finds

\beq
I_2(Q^2) = \frac{2m^2}{Q^2}\int_{0}^{x_0}g_2(x,Q^2)\ dx
=\frac{1}{4}\frac{G_M(Q^2)-G_E(Q^2)}{1+Q^2/4m^2}\,G_M(Q^2)\,,
\label{eq7}
\eeq
i.e. the inelastic contribution for $0 < x <x_0 $ equals the
negative value of the elastic contribution given by the $rhs$ of
Eq.~(\ref{eq7}), which is parametrized  by the magnetic and
electric Sachs form factors $G_M$ and $G_E$, respectively. In the
limit of $Q^2\rightarrow 0$, these form factors are  normalized to
the magnetic moment $\mu_N=\kappa_N+e_N$ and the charge $e_N$ of
the nucleon, $G_M(0)=\mu_N$ and $G_E(0)=e_N$. The BC sum rule has
the limiting cases

\beq
I_2(Q^2)\rightarrow \left \{ \begin{array}{lll}
              \frac{1}{4}\mu_N \kappa_N &{\rm for} &Q^2\rightarrow 0\\
              {\cal O}(Q^{-10})         &{\rm for} &Q^2\rightarrow \infty
         \end{array} \right.\,.
\label{eq8}
\eeq

By use of Eqs.~(\ref{eq5}) the integrals $I_1$ and $I_2$ can be
cast into the form

\beq
I_1(Q^2) = \frac{m^2}{8\pi^2\alpha}\int_{\nu_0}^{\infty}
           \frac{1-x} {1+\gamma^2}
           \left (\sigma_{1/2}-\sigma_{3/2}
           -2\gamma\,\sigma'_{LT}\right )\ \frac{d\nu}{\nu}\ ,
\label{eq9}
\eeq

\beq
I_2(Q^2) = \frac{m^2}{8\pi^2\alpha}\int_{\nu_0}^{\infty}
           \frac{1-x} {1+\gamma^2}
           \left (\sigma_{3/2}-\sigma_{1/2}
           -\frac{2}{\gamma}\,\sigma'_{LT}\right )\ \frac{d\nu}{\nu}\ ,
\label{eq10}
\eeq
where $ \nu_0 =m_{\pi} + (m_{\pi}^2+Q^2)/2m$ is the threshold
$lab$ energy of one-pion production.

Since $\gamma \sigma'_{LT}={\cal O}(Q^2)$, the longitudinal-transverse
term does not contribute to the integral $I_1$ in the real photon
limit.  However, the ratio $\sigma'_{LT}/\gamma$ remains constant in
that limit and hence contributes to $I_2$. As a result we find

\beq
I_2(0) = \frac{1}{4}\kappa_N^2+\frac{1}{4}e_N\kappa_N\ ,
\label{eq11}
\eeq
with the two terms on the $rhs$ corresponding to the contributions
of $\sigma_{3/2}-\sigma_{1/2}$ and $\sigma'_{LT}$, respectively.

It follows from Eqs.~(\ref{eq9}) and (\ref{eq10}) that the sum of
the integrals $I_1$ and $I_2$ is solely determined by the
longitudinal-transverse interference term,

\begin{eqnarray}
\label{eq12} I_3(Q^2)& = &I_1(Q^2)+I_2(Q^2) = \frac{2m^2}{Q^2}
\int_0^{x_0} (g_1+g_2) dx \nonumber \\ & = &
-\frac{m^2}{4\pi^2\alpha} \int_{\nu_0}^{\infty} (1-x) \left
(\frac{\nu}{Q} \sigma'_{LT} \right ) \frac{d\nu}{\nu}\ ,
\end{eqnarray}
with $I_3(0)=\frac{1}{4} e_N\ \kappa_N$. In Eq.~(\ref{eq12}) and in
the following equations we have suppressed the arguments of the spin
structure functions and virtual photoabsorption cross sections,
$(x,Q^2)$ and $(\nu,Q^2)$ respectively. Of course, the integrals of
Eq.~(\ref{eq12}) will only converge if $\sigma'_{LT}$ drops faster
than $1/\nu$ for large values of $\nu$.  The origin of this problem
is, of course, already contained in Eq.~(\ref{eq10}), and there are
severe doubts whether the $I_2$ integral is actually converging.

As we have seen above, $I_1(Q^2)$ approaches the GDH integral in
the limit $Q^2\rightarrow 0$. However, at finite $Q^2$ the
longitudinal-transverse term contributes significantly. In order
to eliminate this term, 3 choices have been made in the
literature~\cite{Pan98}, which in the following will be labeled
$N=A,B$, and $C$,

\begin{eqnarray}
\label{eq13a}
 I_{A} \,(Q^{2}) & = & \frac{m^2}{8\pi^2\alpha}
\int_{\nu_0}^{\infty} (1-x)\ (\sigma_{1/2}-\sigma_{3/2})
\frac{d\nu}{\nu} \nonumber
\\ & = & \frac{2m^2}{Q^2} \int_0^{x_0} (g_1-\gamma^2\
g_2)\ dx\ ,\\ \label{eq13b}
 I_{B} \,(Q^{2}) & = & \frac{m^2}{8\pi^2\alpha}
\int_{\nu_0}^{\infty}
 \frac{1-x}{\sqrt{1+\gamma^2}}\
(\sigma_{1/2}-\sigma_{3/2}) \frac{d\nu}{\nu} \nonumber
\\ & = & \frac{2m^2}{Q^2} \int_0^{x_0}
\frac{1}{\sqrt{1+\gamma^2}}\ (g_1-\gamma^2\ g_2)\ dx\ ,\\
\label{eq13c}
 I_{C} \,(Q^{2}) & = & \frac{m^2}{8\pi^2\alpha}
\int_{\nu_0}^{\infty} (\sigma_{1/2}-\sigma_{3/2}) \frac{d\nu}{\nu}
\nonumber \\ & = & \frac{2m^2}{Q^2} \int_0^{x_0} \frac{1}{1-x}\,
(g_1-\gamma^2\ g_2)\ dx\ .
\end{eqnarray}

It is obvious that definition A is the natural choice in terms of
the quark structure functions. On the other hand, the definitions
B and C look ``natural'' in terms of the transverse-transverse
cross section $\sigma'_{TT}$ if one chooses the definition of the
virtual photon flux according to Gilman or Hand, respectively. As
we shall show in the following section, the numerical differences
between these 3 definitions are quite substantial, in particular
with regard to (I) the slopes at $Q^2=0$ and (II) the positions of
the zero crossing.

Concerning the slopes of the generalized GDH integrals at $Q^2=0$,
we find the following model-independent relations

\begin{eqnarray}
 \label{eq17}
I_B'-I_A'& = & -\frac{m^2}{4\alpha} \frac{1}{4\pi^2}
\int\frac{d\nu}{\nu^3} (\sigma_{1/2}-\sigma_{3/2})\ , \\
\label{eq18} I_C'-I_A' & = &  -\frac{m}{4\alpha} \frac{1}{4\pi^2}
\int\frac{d\nu}{\nu^2} (\sigma_{1/2}-\sigma_{3/2})\ , \\
\label{eq19} I_{1}'-I_A'& = & -\frac{m^2}{2\alpha}
\frac{1}{4\pi^2} \int\frac{d\nu}{\nu^3}
(\sigma_{1/2}-\sigma_{3/2}) -\frac{m^2}{2\alpha} \frac{1}{2\pi^2}
\int\frac{d\nu}{\nu^3} \lim_{Q^2\to 0} \left (\frac{\nu}{Q}
\sigma'_{LT} \right )\ ,
\end{eqnarray}
where $I'=dI/dQ^2$ and all cross sections should be evaluated at
$Q^2=0$. In particular we observe the appearance of the forward
spin polarizability,

\begin{equation}
\gamma_0 = \frac{1}{4\pi^2} \int\frac{d\nu}{\nu^3}
(\sigma_{1/2}-\sigma_{3/2}) \ , \label{eq20}
\end{equation}

and the longitudinal-transverse polarizability

\begin{equation}
\delta_{0} = -\frac{1}{2\pi^2} \int\frac{d\nu}{\nu^3} \lim_{Q^2\to
0} \left (\frac{\nu}{Q} \sigma'_{LT} \right )\ ,
\end{equation}

which can be evaluated pretty safely on the basis of dispersion
relations.

Finally we give the multipole decomposition of the cross sections
for the  dominant one-pion contribution,

\begin{eqnarray}
\label{eq21}
 \sigma'_{TT}(1\pi) & = & 4\pi \frac{k^{cm}_{\pi}}
{K^{cm}} \sum_l\frac{1}{2}(l+1) [-(l+2)(|E_{l+}|^2 +
|M_{l+1,-}|^2)\nonumber
\\ && + l(|M_{l+}|^2 + |E_{l+1,-}|^2) - 2l(l+2)
(E_{l+}^{\ast}M_{l+}-
     E^{\ast}_{l+1,-}M_{l+1,-})] \nonumber \\
& = & 4\pi \frac{k^{cm}_{\pi}} {K^{cm}}\{-|E_{0+}|^2 + |M_{1+}|^2
- 6E_{1+}^{\ast}M_{1+} - 3|E_{1+}|^2 + |E_{2-}|^2 \pm\ ... \}\,
\\
 \label{eq22}
 \sigma'_{LT}(1\pi) & = & 4\pi \frac{k^{cm}_{\pi}} {K^{cm}}
 \left (\frac{Q}{\omega^{cm}_{\gamma}}\right
  )  \sum_l\frac{1}{2}(l+1)^2 \nonumber \\
&& \cdot \  [-L_{l+}^{\ast}((l+2)E_{l+} + lM_{l+}) +
L_{l+1,-}^{\ast}
   (lE_{l+1,-} + (l+2)M_{l+1,-})] \nonumber \\
& = & 4\pi \frac{k^{cm}_{\pi}} {K^{cm}}\left
(\frac{Q}{\omega^{cm}_{\gamma}}\right
  )\{-L_{0+}^{\ast}E_{0+} - 2L_{1+}^{\ast}(M_{1+}+3E_{1+}) \nonumber \\
&&  + L_{1-}^{\ast}M_{1-} + 2L_{2-}^{\ast}E_{2-}\pm\ ...\}\ .
\end{eqnarray}

We note that here and in the following, expressions as
$E_{l+}^{\ast}M_{l+}$ should be read as Re$(E_{l+}^{\ast}M_{l+})$.
 The kinematical variables in Eqs.~(\ref{eq21}) and~(\ref{eq22})
  are $K^{cm} = \frac{W^{2}
- m^{2}}{2W} ,\ \omega^{cm}_{\gamma} = \frac{m\nu-Q^2}{W}$ and
$k^{cm}_{\pi} = \frac{1}{2W}
\sqrt{(W^2-m_{\pi}^2)^2-2m^2(W^2+m_{\pi}^2)+m^4}$, where
$W=\sqrt{m^2+2m\nu-Q^2}$ is the total $c.m.$ energy. Since
$L_{0+}$ has a zero at $\omega^{cm}_{\gamma}=0$, it is often
convenient to use the Coulomb or ``scalar'' multipoles defined by

\begin{equation}
S_{l\pm} = \frac{k^{cm}_{\gamma}}{\omega^{cm}_{\gamma}}\ L_{l\pm}
\ , \ \label{eq23}
\end{equation}

with $ k^{cm}_{\gamma} = \frac{m}{W} \sqrt{\nu^{2}+Q^{2}}\ $ .

In particular Eq.~(\ref{eq19}) can be cast into the form

\begin{equation}
I_1'-I_A' = -\frac{m^2}{2\alpha} (\gamma_0-\delta_0)\ ,
\label{eq24}
\end{equation}
with the forward spin polarizability

\begin{equation}
\gamma_0 = \frac{2}{\pi} \int \frac{d\nu}{\nu^3}
\frac{k_{\pi}^{cm}}{\nu} \sqrt{1+\frac{2\nu}{m}} \{\mid
E_{0+}\mid^2 - \mid M_{1+}\mid^2 + 6E_{1+}^{\ast}M_{1+} + 3\mid
E_{1+}\mid^2 \pm\ ...\}
\label{eq25}
\end{equation}
and the longitudinal-transverse polarizability

\begin{equation}
\delta_0 = \frac{2}{\pi} \int \frac{d\nu}{\nu^3}
\frac{k_{\pi}^{cm}}{\nu} \left (1+\frac{2\nu}{m}\right )
\{L_{0+}^{\ast}E_{0+}+2L_{1+}^{\ast}(M_{1+}+3E_{1+}) \pm\ ... \}\
. \label{eq26}
\end{equation}

Due to the weight factor $\nu^{-3}$ in the integral for
$\gamma_0$, Eq.~(\ref{eq25}), the contribution of s-wave pion
production $(E_{0+})$ is enhanced such that it nearly cancels the
contribution of magnetic $\Delta$ excitation $(M_{1+})$. Though
the electric quadrupole excitation is small, $E_{1+}/M_{1+}\approx
-2.5\%$ for real photons, its interference term with the magnetic
dipole excitation becomes quite important for the forward spin
polarizability because of the mentioned cancellation. However,
such a cancellation does not occur in the case of $\delta_0$,
Eq.~(\ref{eq26}). The Fermi-Watson theorem asserts that
$L_{l\pm},\ E_{l\pm}$ and $M_{l\pm}$ have the same phases in the
region of interest below two-pion threshold, and gauge invariance
requires that $E_{l\pm}=L_{l\pm}$ in the Siegert limit or ``pseudo
threshold'' situated in the unphysical region. The latter relation
changes in the physical region, and as a rule of thumb we find
$L_{0+}\approx\frac{1}{2}E_{0+}$ and $L_{1+}\approx 2E_{1+}$ in
the (physical) threshold region. Since the p-wave contribution in
Eq.~(\ref{eq26}) is much suppressed, there is no substantial
cancellation between s and p waves, and as a result the
longitudinal-transverse term gives the dominant contribution to
the difference of slopes between $I_1$ and $I_A$, e.g. in
Eq.~(\ref{eq24}).

\section{Results and Discussion}

\begin{figure}[h]
\centerline{\psfig{file=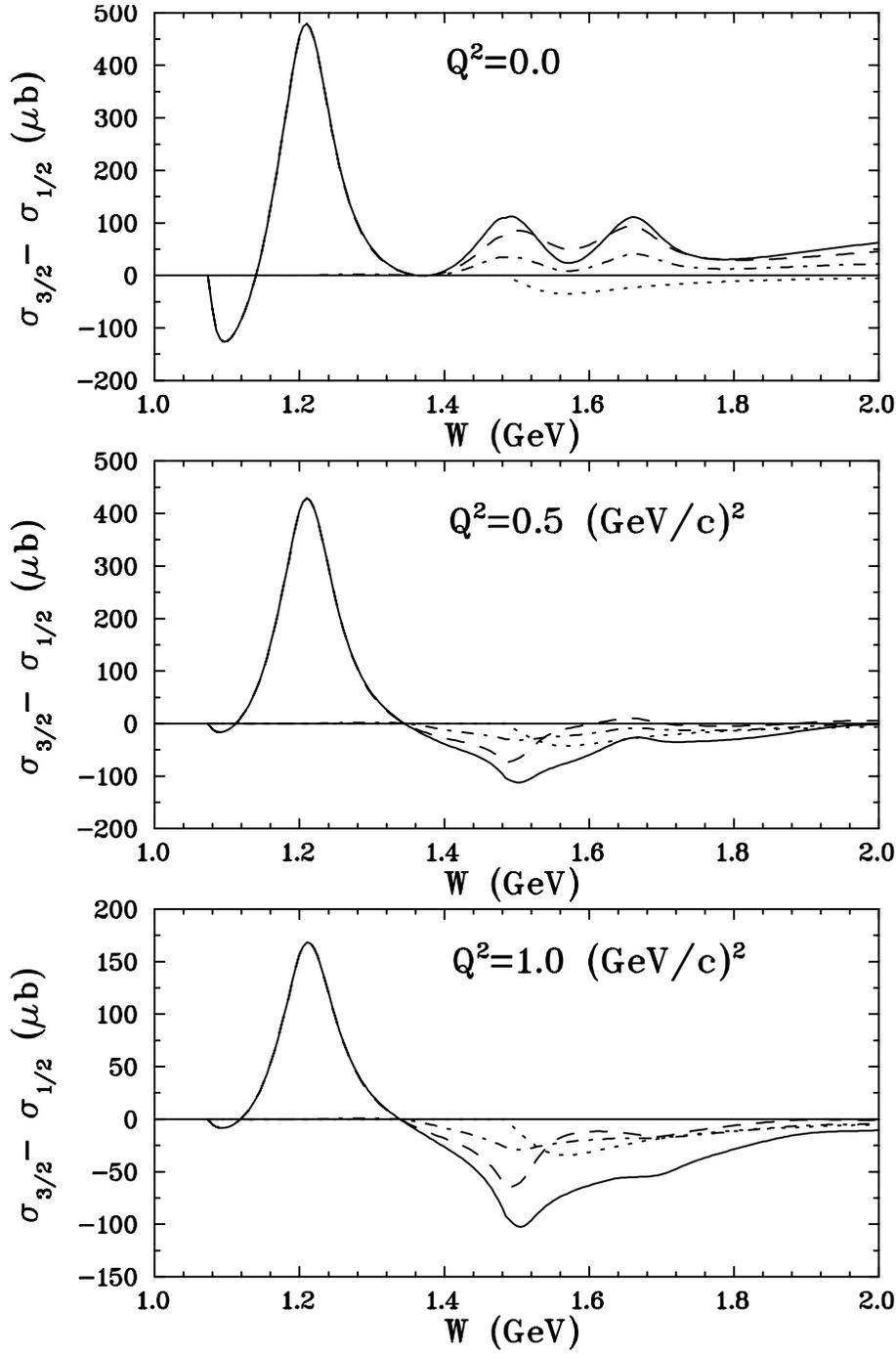,width=12cm,silent=}}
\vspace{0.5cm}
 \caption
 { \label{fig1} Helicity difference $\sigma_{3/2} - \sigma_{1/2}$ for
   the proton at $Q^{2} = 0,\,\, 0.5$ and 1.0~(GeV/c)$^{2}$ Dashed,
   dotted and dash-dotted curves: contributions of single-pion, eta
   and multipion channels, respectively; solid curves: total result.}
\end{figure}

\begin{figure}[h]
\centerline{\psfig{file=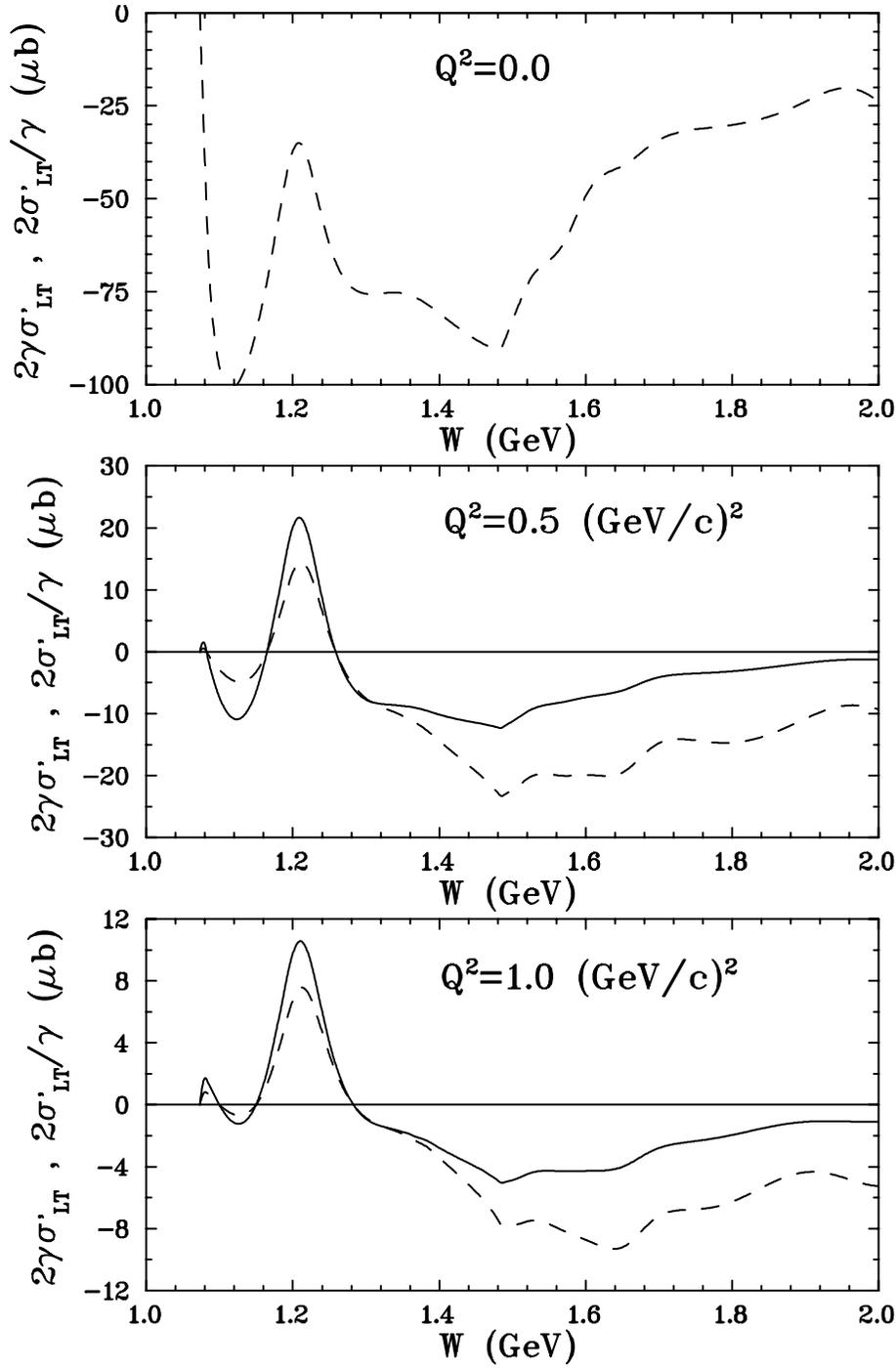,width=12cm,silent=}}
\vspace{0.5cm}
\caption
{
\label{fig2} Longitudinal-transverse cross sections for the proton,
  $2\gamma\sigma^{'}_{LT}$ (solid curves) and
  $2\sigma^{'}_{LT}/\gamma$ (dashed curves) at $Q^{2} = 0,\,\, 0.5$
  and 1.0~(GeV/c)$^{2}$. The calculations are for the one-pion
  channel only.
}
\end{figure}

The spin structure of the nucleon is described by the cross
sections $ \sigma_{3/2}\, -\, \sigma_{1/2}$ and $\sigma'_{LT}$,
which are plotted in Figs.~\ref{fig1} and~\ref{fig2} as functions
of the total energy $W$ for $Q^{2} = 0, 0.5$ and $1.0\,
(GeV/c)^{2} $. In the upper part of Fig.~\ref{fig1}, at $Q^2=0$,
the negative contribution near threshold is essentially due to
s-wave production of charged pions. Due to the weighting with
$\nu^{-1}$ and $\nu^{-3}$, for the GDH integral and the spin
polarizability respectively, s-wave pions become increasingly
important in comparison with p-wave production near the $ \Delta
(1232) $, which carries the opposite sign. At the photon point,
$Q^{2} = 0$, the second and third resonance regions are clearly
visible, their contributions have the same sign as for the
$\Delta$.

With increasing momentum transfer $Q^{2}$, the cross sections
generally decrease as is to be expected. However, there are two
interesting peculiarities, which are of some consequence for the
$Q^{2}$ dependence of the generalized GDH integrals.

\begin{enumerate}
\item[(I)]
The negative bump due to s-wave pion production near threshold
decreases rapidly with increasing values of $Q^{2}$ due to the
form factor. In the case of the $\Delta$, this effect is mostly
compensated by the increase of the p-wave multipole appearing with
the 3-momentum of the virtual photon. As a result the $\Delta$
effects become rapidly more dominant at very small $ Q^{2}$.
\item [(II)]
The contributions in the second and third resonance region, which
added to the $\Delta$ contribution at $Q^{2} = 0$, change sign at
$Q^{2} \approx 0.5\,(GeV/c)^{2}$.
\end{enumerate}

The general decrease due to form factors and the change of sign in
the first vs. second and third resonance regions lead to a rapid
decrease of the integrals in absolute values.

The second structure function $\sigma_{LT}^{'}$ is shown in
Fig.~\ref{fig2}. This longitudinal-transverse cross section
vanishes, of course, for real photons. However, it contributes to
the Burkhardt-Cottingham sum rule even in the limit
$Q^{2}\,\to\,0$ due to Eq.~(\ref{eq10}), where it appears
multiplied by a factor $\nu/Q$ in the integral. In Fig.~\ref{fig2}
we therefore show both $2\gamma \sigma^{'}_{LT}$ and
$2\sigma^{'}_{LT}/\gamma$, which contribute to the $I_{1}$ and
$I_{2}$ integrals, respectively. Obviously the convergence of
$2\sigma^{'}_{LT}/\gamma$ is bad, even after the weighting with
$\nu^{-1}$ in Eq.~(\ref{eq10}). The different sign of s-wave and
$\Delta$ contributions is also seen in Fig.~\ref{fig2}. Since
$L_{1+}/M_{1+}$ is of the order of $5\%$ for real photons, the
$\Delta$ resonance appears in that case as a small bump on a large
negative background. Again the s-wave threshold contribution
decreases rapidly with increasing $Q^2$, such that the $\Delta$
resonance becomes the dominant feature for $Q^2=1$~(GeV/c)$^2$ and
$W<1.3$~GeV. Due to the increasingly negative contribution for
larger energies, however, the integral over the
longitudinal-transverse cross sections of Fig.~\ref{fig2} will
always take the negative sign, such that the contribution of
$\sigma'_{LT}$ to $I_1$ and $I_2$ (see Eqs.~(\ref{eq9})
and~(\ref{eq10})) will be positive at every value of $Q^2$.

\begin{figure}[h]
  \centerline{\psfig{file=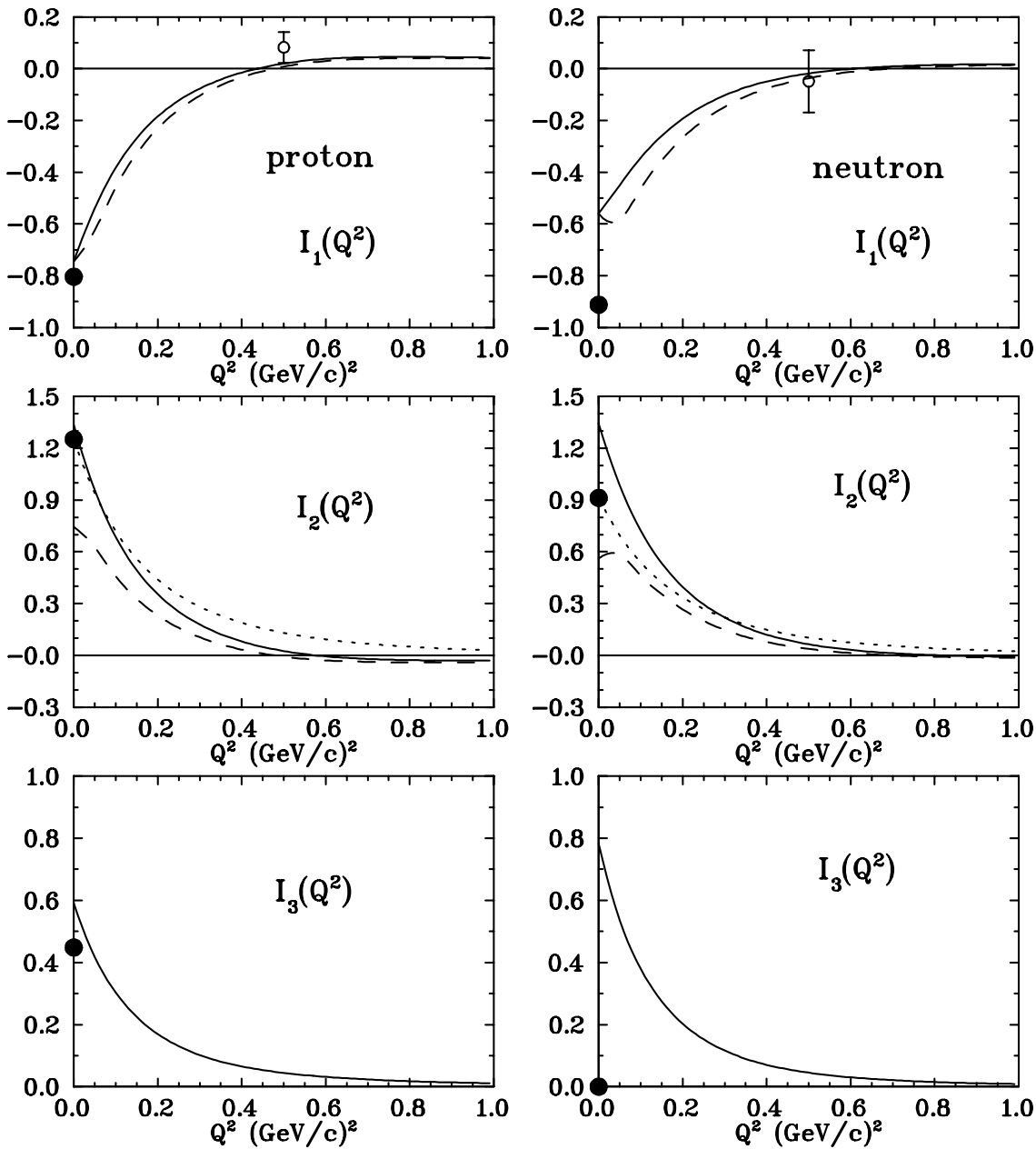,width=15cm,silent=}}
\vspace{0.5cm}
  \caption
  {\label{fig3} The integrals $I_1$, $I_2$ and $I_{3}$ for the protons
    and neutrons as functions of $Q^{2}$ and integrated up to $W_{max}
    = 2$~GeV. Dashed curves are the results obtained without the
    $\sigma'_{LT}$ contribution, solid curves are the total results,
    and dotted curves for the $I_{2}$ integrals are the prediction of
    the BC sum rule. The full circles at $Q^{2} = 0$ are the GDH sum
    rule values. The open circles at $Q^{2} = 0.5$~(GeV/c)$^{2}$ are
    the recent SLAC data~\protect\cite{Abe98} for the resonance region. }
\end{figure}

\begin{figure}[h]
  \centerline{\psfig{file=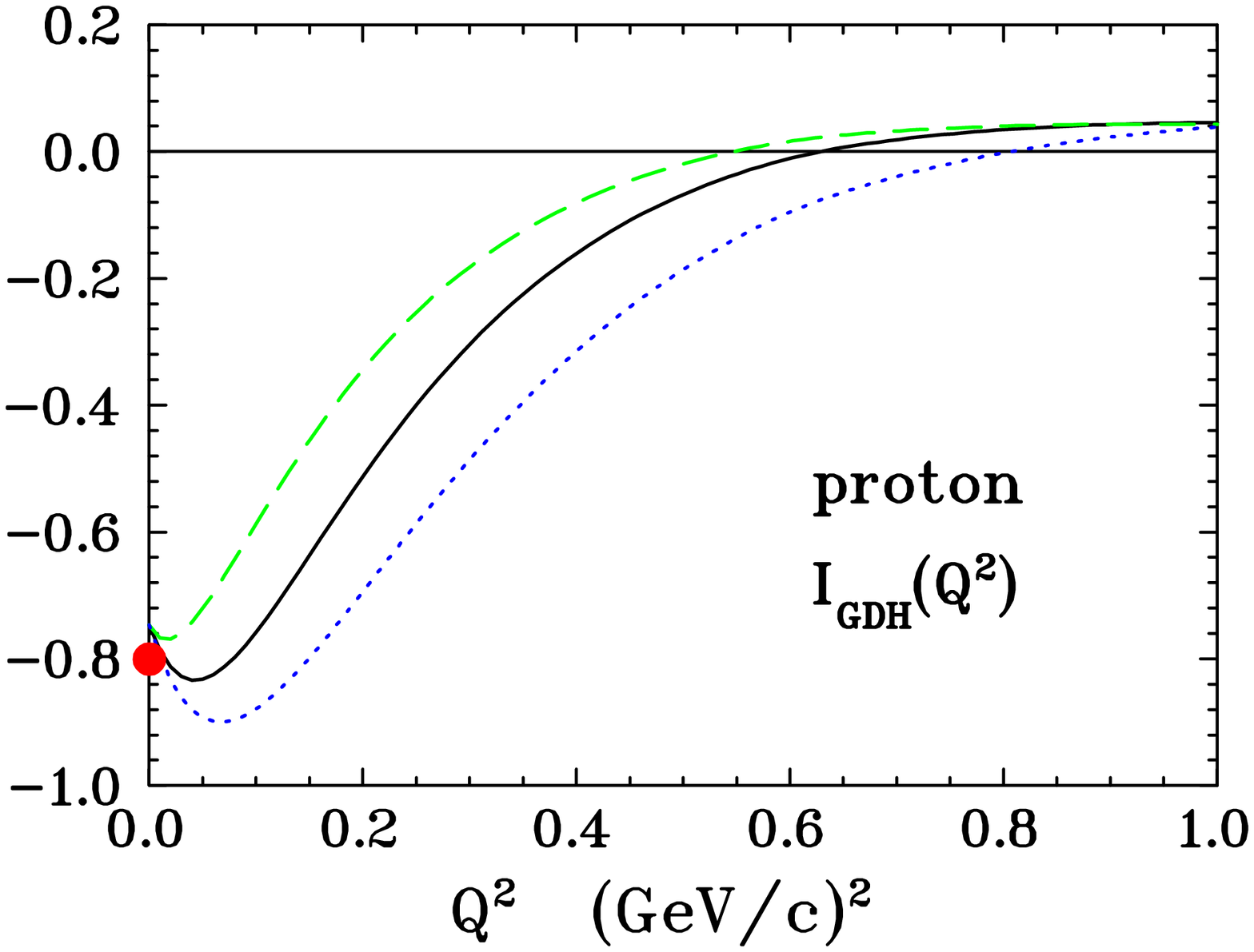,width=6cm, angle=90, silent=}
  \hspace{0.3cm}
 \psfig{file=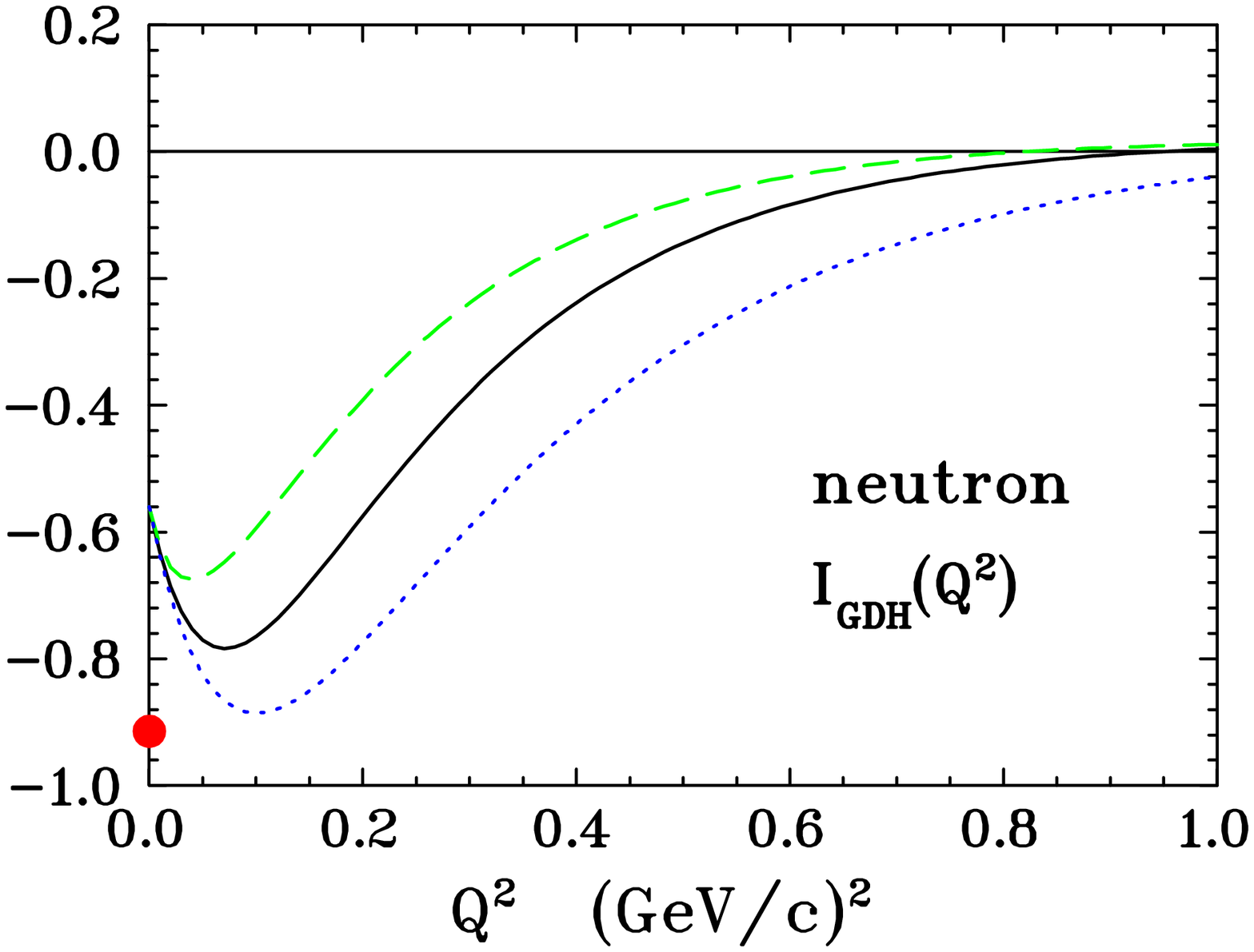,width=6cm, angle=90,  silent=}}
  \vspace{0.5cm} \caption
{\label{fig4} The GDH integrals $I_A$ (solid
    curves), $I_B$ (dashed curves) and $I_C$ (dotted curves) for
    proton and neutron, integrated up to
    $W_{max} = 2$~GeV, in the range $0\le Q^2\le 1$~(GeV/c)$^2$.
     The full circles at $Q^{2} = 0$ are the GDH
    sum rule values.}
\end{figure}

\begin{figure}[h]
  \centerline{\psfig{file=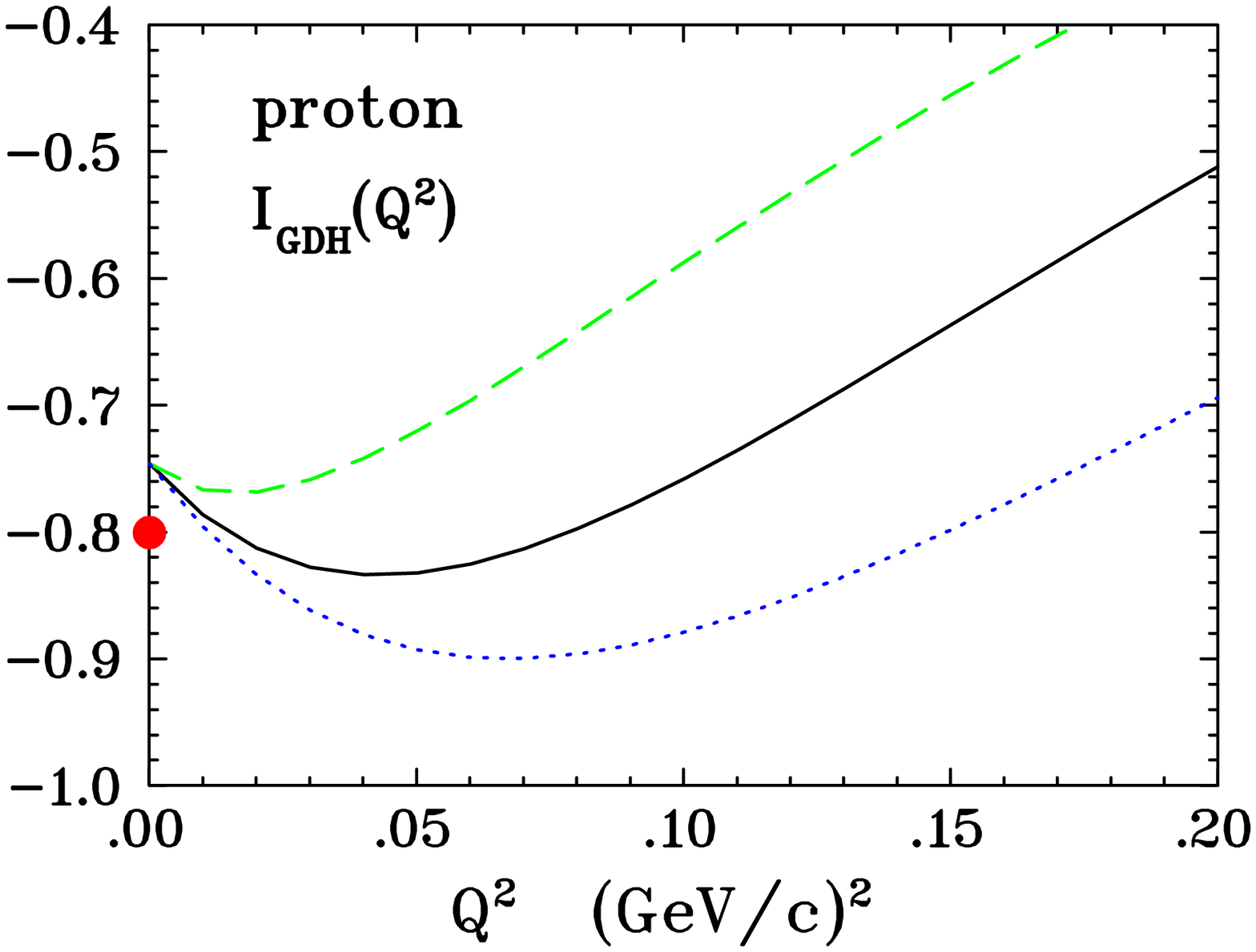,width=6cm, angle=90,  silent=}
\hspace{0.3cm}
 \psfig{file=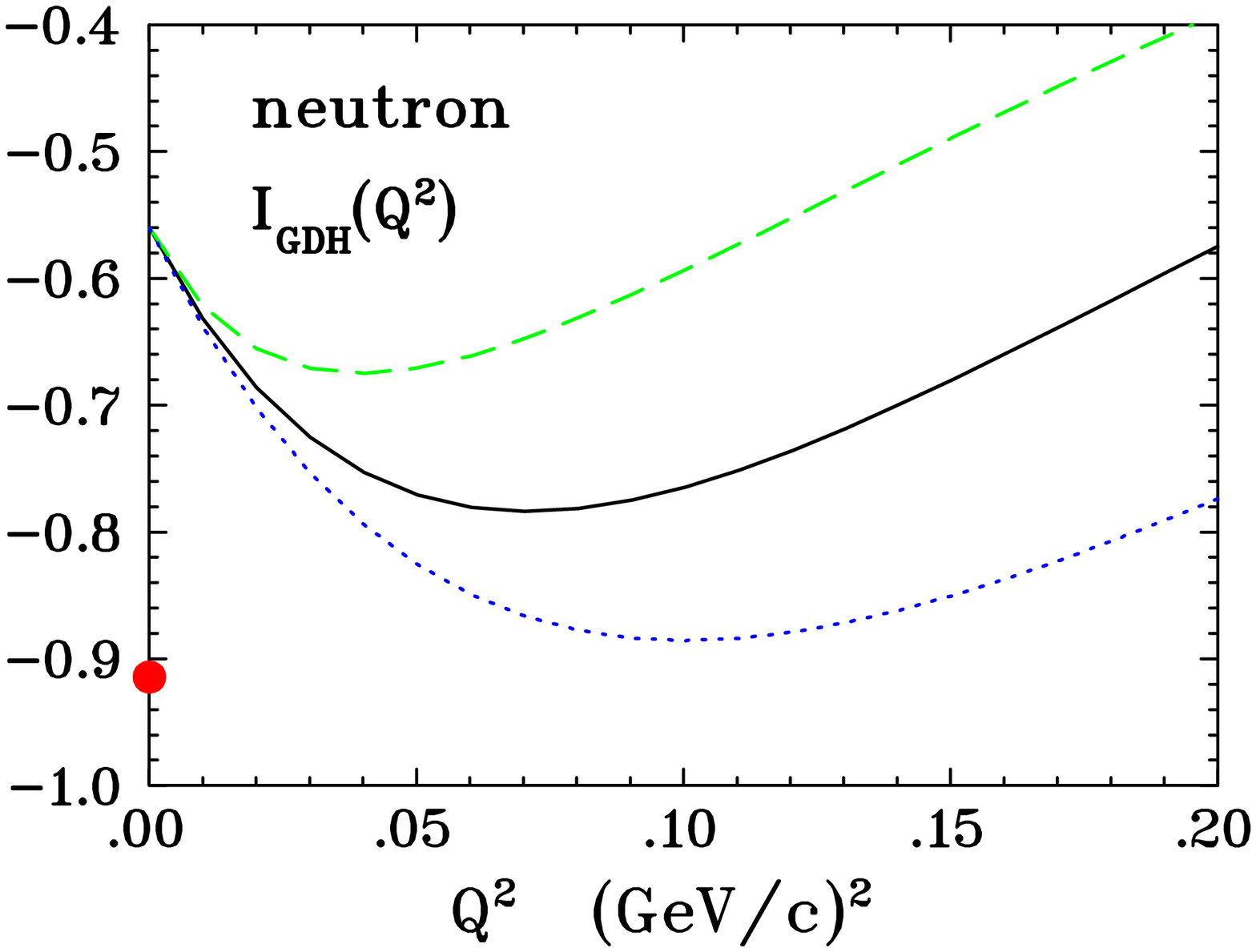,width=6cm, angle=90,  silent=}}
\vspace{0.5cm} \caption {\label{fig5} The GDH integrals $I_A,\
I_B,\ $ and $I_C,$ at small $Q^2$. Notation as in
Fig.~\protect\ref{fig4}.}
\end{figure}

The dependence of various generalized GDH integrals on momentum
transfer is shown in Figs.~\ref{fig3}-\ref{fig5}. We include the
one-pion channel according to MAID~\cite{Dre99a} and the two-pion
and eta channels as in our previous work~\cite{Dre99b}. It is
obvious that the integrals $I_A,\ I_B$, and $I_C$ with purely
transverse cross sections in the integrand (see Fig.~\ref{fig4})
behave fundamentally different from $I_{1}$ (see Fig.~\ref{fig3}),
which also gets contributions from the longitudinal-transverse
interference. While the latter has a large positive slope at the
origin and crosses the zero line around $Q^{2} = 0.4
\,(GeV/c)^{2}$, the former 3 integrals have negative slopes at the
origin, which leads to zero-crossings at much larger $Q^{2}$. The
minima of the generalized GDH integrals occur at very small
momentum transfers, indeed, as may be seen in more detail in
Fig.~\ref{fig5}. Such negative slopes at the origin were first
obtained by Burkert, Ioffe and others~\cite{Bur92,Azn95} in the
framework of phenomenological models. In a similar spirit Scholten
and Korchin~\cite{Sch99} have recently evaluated $I_A(Q^2)$ in the
framework of an effective Lagrangian model. Their results are in
qualitative agreement with those of Fig.~5, in particular they
also predict a minimum at $Q^2=0.5\,(GeV/c)^2$. However, it is
worthwhile pointing out that the different definitions lead to
quite different results amongst each other. It is therefore
absolutely prerogative to give clear definitions before comparing
the published experimental or theoretical values. Concerning such
definitions we find it most convenient to express the integrals in
terms of the spin structure functions $g_{1}$\, and $g_{2}$,
because these are uniquely defined in the literature.

As mentioned in section 1, there have been many phenomenological
estimates for the GDH sum rule of the nucleon. Until recently the
estimates for the GDH integral of the proton were in the range
from -260~$\mu$b to -290~$\mu$b, while the sum rule predicted
-204~$\mu$b~\cite{Dre98}. Our present analysis is shown in
Table~\ref{tab1}. In the region 200~MeV$<\nu<$800~MeV we can now
rely on the experimental data taken in 1998 at MAMI~\cite{Ahr00}.
We are quite confident about our estimate for the integrand
between threshold and 200~MeV, because here the cross sections are
largely determined by s-wave pion production via the amplitude
$E_{0+}$ fixed through low energy theorems. Also our estimate for
one-pion production in the energy range 800~MeV$<\nu<$ 1.6~GeV
should be quite reliable, because our model MAID describes the
helicity structure of the main resonances very well. Since $\eta$
production has now been measured over a large energy range, and
because of its $S_{11}$ dominance, the helicity structure of that
contribution is fixed. The situation is less clear with regard to
$K^{+}$ production, but the overall contribution of this reaction
should be small. The only major uncertainty is with regard to
two-pion production at the higher energies, which has been
estimated with the prescription of Ref.~\cite{Van00b}.

As can be seen from Table~\ref{tab1}, the contributions of
one-pion production from below $200~$MeV and above $800~$MeV
cancel almost. The still missing DIS contribution has been
estimated by Bianchi et al.~\cite{Bia99} from an extrapolation of
DIS to real photons. If we add those $(26 \pm 7)~\mu$b to our
analysis in the resonance region, the total result for the proton
is $(-202 \pm 10)~\mu$b. While the agreement between analysis and
sum rule is remarkably good for the proton, this does not yet end
the story. In fact our present prediction for the neutron is off
the sum rule value by about $60~\mu$b, i.e. $25\%$!

The slope of the integral $I_C$ was first obtained by Bernard et
al.~\cite{Ber93} in chiral perturbation theory (ChPT) to $O (p^{3})$,
with the result

\begin{equation}
 I_C' = \frac{1}{6} m^2 \left ( \frac{g_{A}}{4\pi
m_{\pi} f_{\pi}} \right )^2\ . \label{eq26a}
\end{equation}

Recently, Ji et al.~\cite{JiO99} have calculated the slope of the
integral $I_A$ up to terms of $O(p^{4})$, with the result

\begin{equation}
  I_A' = \frac{1}{6} m^2 \left (\frac{g_{A}}{4\pi m_{\pi} f_{\pi}}
  \right ) ^2 \Bigg\lbrace 1 - \frac{\pi}{4} \frac{m_{\pi}}{m} (13 + 2
  \tau_{3} + 2 \kappa_{V})\Bigg\rbrace \ .
 \label{eq27}
\end{equation}

By comparing Eqs.~(\ref{eq26a}) and~(\ref{eq27}) we conclude that
$I_A'$ and $I_C'$ take the same value at $O (p^3)$. The reason for
this becomes obvious from the model-independent relations
Eqs.~(\ref{eq17}) -~(\ref{eq19}). While the $rhs$ of
Eqs.~(\ref{eq17}) and~(\ref{eq19}) is $O(m^2/m_{\pi}^2)$, as
follows explicitly from the ChPT result for the forward spin
polarizability $\gamma_0$, the $rhs$ of Eq.~(\ref{eq18}) is
$O(m/m_{\pi})$, i.e. one order higher in the chiral counting
scheme. Clearly the $(m/m_{\pi})^2$ or $(m/m_{\pi})$ behavior of
the $rhs$ is due to the fact that the integrals diverge like
$m_{\pi}^{-2}$ or $m_{\pi}^{-1}$, respectively, if the lower limit
of the integrals, $\nu_0\simeq m_{\pi}$, goes to zero.

The slope of $I_{1}$, on the other hand, vanishes to $O (p^{3})$.
To next order the result is~\cite{JiO99}

\begin{equation}
I_1' = \frac{1}{6}m^2 \left(\frac{g_{A}}{4{\pi}
m_{\pi}f_{\pi}}\right)^{2} \frac{\pi}{8} \, \frac{m_{\pi}}{m}
\,(1+3 \kappa_{V}+2\tau_{3}+ 6\kappa_{S} \tau_{3})\ . \label{eq28}
\end{equation}

The forward spin polarizability was first calculated to $
O(p^{3})$ by Bernard et al.~\cite{Ber95}. Again large corrections
were found in the next order, $O(p^{4})$, by Ji et
al.~\cite{JiK99} and Kumar et al.~\cite{Kum99}. However, Gellas et
al.~\cite{Gel00} claim that these authors used a definition of
$\gamma_{0}$ that differed from the usual definition through
forward dispersion relations as in Eq.~(\ref{eq25}), i.e. that the
spin polarizabilities of Refs.~\cite{JiK99,Kum99} contain higher
order contributions from the nucleon pole term. The origin of the
problem is due to the fact that HBChPT relies on nonrelativistic
expansions of the intermediate state propagators, which makes it
difficult to separate the pole terms from the internal structure
contributions that are related to the polarizabilities. The two
results for $\gamma_{0}$ at $O(p^{4})$
read~\cite{Gel00,JiK99,Kum99}

\begin{equation}
\gamma_{0}\ {\rm{(GHM)}} =
\frac{2}{3}\alpha\left(\frac{g_{A}}{4\pi \,
m_{\pi}f_{\pi}}\right)^{2}\, \Bigg\{1- \frac{\pi}{8} \frac{
m_{\pi}}{m}\,(13+4\tau_{3}+\kappa_{V}-\kappa_{S}\tau_3)\Bigg\}\ ,
\label{eq29}
\end{equation}

\begin{equation}
\gamma_{0}\ {\rm{(JKO,\ KMB)}} =
\frac{2}{3}\alpha\left(\frac{g_{A}}{4\pi \,
m_{\pi}f_{\pi}}\right)^{2}\, \Bigg\{1- \frac{\pi}{8} \frac{
m_{\pi}}{m}\,(15+6\tau_{3}+3\kappa_{V}+\kappa_{S}\tau_3)\Bigg\}\ .
\label{eq29a}
\end{equation}

The value of $\delta_{0}$ can then be obtained by the
model-independent relation Eq.~(\ref{eq24}). If we assume that the
results for $I_A'$ and $I_1'$ of Ref.~\cite{JiO99} are valid,
Eqs.~(\ref{eq27})-(\ref{eq29a}) lead to the expressions

\begin{equation}
\delta_{0}\ {\rm{(GHM)}} =
\frac{1}{3}\alpha\left(\frac{g_{A}}{4\pi \,
m_{\pi}f_{\pi}}\right)^{2}\,
\Bigg\{1+\frac{\pi}{8}\frac{m_{\pi}}{m}\,(1-2\tau_{3}
+5\kappa_{V}+8\kappa_{S}\tau_{3})\Bigg\}\ ,\label{eq30}
\end{equation}

\begin{equation}
\delta_{0}\ {\rm{(JKO,\ KMB)}} =
\frac{1}{3}\alpha\left(\frac{g_{A}}{4\pi \,
m_{\pi}f_{\pi}}\right)^{2}\,
\Bigg\{1-\frac{\pi}{8}\frac{m_{\pi}}{m}\,(3+6\tau_{3}
-\kappa_{V}-4\kappa_{S}\tau_{3})\Bigg\}\ . \label{eq30a}
\end{equation}

The values for the polarizabilities of the proton are compared to
our results in table~\ref{tab2}. Here and in the following all
numerical expressions are obtained with $\kappa_{V} = 3.71 $ and $
\kappa_{S} = -0.12$ the isovector and isoscalar combinations of
the anomalous magnetic moments, $g_{A} = 1.26$ the axial coupling
constant, $f_{\pi} = $92.4~MeV the pion decay constant, $m=$
938~MeV the proton mass, and $m_{\pi} = $138~MeV an average pion
mass.

The upper part of table~\ref{tab2} shows a partial wave
decomposition of the polarizabilities. As has been pointed out
before, the contributions of s and p waves nearly cancel in the
case of $\gamma_0$, while $\delta_0$ is largely dominated by s
waves. In the central part of table~\ref{tab2}, we give the
contributions of various energy bins. Due to the weighting factor
$\nu^{-3}$, the threshold region below 200~MeV contributes
substantially, while the region above 800~MeV is of little
importance. In the case of $\gamma_0$, the integrand in the
resonance region (200~MeV$<\nu<$800~MeV) has now been measured at
MAMI~\cite{Ahr00}. The experimental value is
(-1.71$\pm$0.09)~$10^{-4}$~fm$^4$, slightly smaller than the
prediction of MAID, -1.66$\cdot 10^{-4}$~fm$^4$. We conclude that
the presently best value of the forward spin polarizability is
$\gamma_0=(-0.70\pm0.10)\,10^{-4}~$fm$^4$. Comparing now with the
results of ChPT at $O(p^3)$, we find the interesting relation
$\gamma_0=2\delta_0$, which can be explained as follows. The
leading order term originates from the lower limit of the
integrals in Eqs.~(\ref{eq19}) and~(\ref{eq20}) if $\nu_0\simeq
m_{\pi}\rightarrow 0$. In that limit the threshold
amplitudes~\cite{Sch91} for charged pion production are
$E_{0^+}^{\rm{thr}} (\pi^+) = 2L_{0^+}^{\rm{thr}}(\pi^+)$. By
comparing Eqs.~(\ref{eq25}) and~(\ref{eq26}) in the same limit,
$\nu$ and $m_{\pi}\ll m$, all higher partial waves vanish and we
are left with integrands $\nu^{-2}$ times $\mid E_{0^+}\mid^2$ and
$L_{0^+}^{\ast}E_{0^+}$, respectively, which immediately leads to
the mentioned relation. Given the existing ambiguity in the
definition of $\gamma_0$ in HBChPT , it is probably not very
useful to compare the phenomenological results with the
predictions of ChPT. A brief look at the last 4 lines of
table~\ref{tab2} shows, however, that neither prediction can
presently describe both $\gamma_0$ and $\delta_0$. It would
therefore be interesting to examine whether the higher order terms
in $I_1'$ etc. could possibly also be affected by the ambiguities
to separate elastic and inelastic contributions.

In table~\ref{tab3} we compare our predictions for the slopes of
the integrals of Eqs.~(\ref{eq9})-(\ref{eq13c}) with the
predictions of ChPT, the Burkhardt-Cottingham sum rule, and the
models of Anselmino et al., and Soffer and Teryaev. We should
point out that we have evaluated all predictions with the same set
of coupling constants and masses as specified before. Furthermore,
we have used the more recent value $\Gamma_1^p=0.146$ for the
asymptotic value~\cite{Glu96} of the integral over the spin
structure function $g_1$, and the proton radii $\langle
r^2\rangle_M^p=0.728$~fm$^2$ and $\langle
r^2\rangle_E^p=0.717$~fm$^2$ as given in Ref.~\cite{Mer96}.

It is obvious that the different approaches lead to rather
different results. In particular, the generalized GDH integrals
differ by factors and, with regard to ChPT at $O(p^{3})$, even by
sign. In the case of $I_B$ we can estimate the asymptotic
contribution, which is missing in our analysis, from the HERMES
data~\cite{Ack98}. If we add the slope of this contribution to our
value, we obtain a further decrease by about $20\%$, i.e. a final
result of $I_B' \approx -4$~(GeV/c)$^{-2}$, which is still
different from ChPT by a factor of 3. As has been pointed out
before, the $O(p^3)$ calculation predicts $I_A'=I_C'$ and
$I_1'=0$. If we use our model-independent Eqs.~(\ref{eq17})
and~(\ref{eq24}) and the relation $\gamma_0=2\delta_0$ at
$O(p^3)$, we find $I_B'=0$. In a similar way we have evaluated
$I_B'=-6.6$~GeV$^{-2}$ at $O(p^4)$ from the results of Ji et
al.~\cite{JiO99,JiK99}. Had we taken the value of $\gamma_0$ from
Ref.~\cite{Gel00}, the result would be $I_B'=-12.6$~GeV$^{-2}$. As
a consequence the ambiguity in the $O(p^4)$ spin polarizabilities
influences the predictions for the generalized GDH integrals and a
solution of the problem is urgently called for. We recall that
this problem arises from the difficulty to describe the proper
pole structure for real and virtual Compton scattering in higher
order HBChPT, and repeat our suspicion that similar ambiguities
could also arise in the case of the GDH integrals.

We further note that our analysis does not support the conjecture
of Soffer and Teryaev~\cite{Sof93} that $I_3=I_1+I_2$ varies
slowly with $Q^{2}$. These authors propose to parametrize $I_3$
for the proton by a smooth function interpolating between the
given values for $Q^2=0$ and $Q^2\rightarrow \infty$,

\begin{equation}
\label{eq34}
I_3\, (Q^2) = \left\{ \begin{array}{r@{\quad,\quad}l}
              \kappa_p/4 -2m^2\, Q^2\, \Gamma^p_1/Q_0^4 &
              Q^2<Q_0^2 \\
              2m^2\, \Gamma^p_1 /Q^2 & Q^2>Q_0^2 \ .
              \end{array} \right.
\end{equation}

The continuity of the function and its derivative is implemented
by choosing $Q_0^2=16m^2\, \Gamma^p_1/\kappa_p \simeq
1$~(GeV/c)$^2$. The resulting derivative at the origin is very
small,

\begin{equation}
I_3'= -\frac{\kappa^2_p}{128m^2\, \Gamma^p_1} \approx
-0.2~{\rm{GeV}}^{-2} \ ,\label{eq35}
\end{equation}

and the derivative of $I_1$ follows from assuming the validity of
the BC sum rule. In comparing our result with Eq.~(\ref{eq35}), we
find that the slopes differ by about a factor of 20.

The model of Anselmino et al.~\cite{Ans89}, on the other hand, is
relatively close to our prediction. Motivated by the vector
dominance model, these authors proposed to describe $I_1$ by the
expression

\begin{equation}
I_1\, (Q^2) =  \frac{1}{(Q^2+m_V^2)^2} \left
              (2m\Gamma^p_1\, Q^2\, -\frac{\kappa^2_p}{4}\, m_V^4\right )\ ,
\label{eq36}
\end{equation}

with $m_V\approx m_{\rho}$, the mass of the $\rho$ meson. The
resulting derivative at the origin is

\begin{equation}
I_1' = \frac{\kappa^2_p}{2m_V^2} +
              \frac{2m^2\Gamma^p_1}{m_V^4} \simeq 3.4~{\rm{GeV}}^{-2}\,
              \label{eq37}
\end{equation}

while our phenomenological model predicts 4.4~GeV$^{-2}$.

\section{SUMMARY}
The results of our phenomenological model MAID have been presented for
various spin observables. We have updated our prediction for the GDH
sum rule for the proton and find a good agreement with the sum rule
value. The change from earlier results originate from a collaborative
effect of a correct treatment of the threshold region, a reduced
estimate for the two-pion contributions based on the recent results of
the GDH Collaboration at MAMI, and an asymptotic contribution. We look
forward to the continuation of the MAMI experiments to the higher
energy at ELSA, which will probe the weakest point of our prediction,
the two-pion contribution in the resonance region. In the case of the
neutron the disagreement remains, and we are urgently waiting for a
direct measurement of the helicity-dependent neutron cross sections in
the resonance region.

The same model can be used to calculate the spin polarizability and an
associated longitudinal-transverse polarizability. Due to the
weighting factor $\nu^{-3}$, these predictions are on rather firm
ground. Unfortunately, a comparison with the results from ChPT is
difficult, because their interpretation is still under discussion.

We have also presented a systematic analysis of various generalized
GDH-type integrals $I_{\rm{GDH}}(Q^2)$ and of the BC sum rule
$I_2(Q^2)$. In the case of the proton, the latter can be reasonably
well described by our model, but our prediction for the neutron fails
again. Though all generalized GDH integrals are fixed for both real
photons, $Q^2=0$, and in the asymptotic region,
$Q^2\rightarrow\infty$, their dependence on momentum transfer differs
considerably for $Q^2\lesssim 1$~(GeV/c)$^2$. A clear signature of
these differences are the slopes $I_{\rm{GDH}}'=dI_{\rm{GDH}}/dQ^2$ at
the origin. In particular we compare our predictions to those of other
phenomenological models and heavy baryon chiral perturbation theory.
The different slopes from different definitions of $I_{\rm{GDH}}$ are
related by model-independent relations involving the spin
polarizability and related quantities, which can be safely evaluated
from the helicity-dependent cross sections in the resonance region.
Since both the spin polarizability and the derivatives of the GDH
integrals are derived from doubly virtual Compton scattering in chiral
perturbation theory, the ongoing discussion on how to separate pole
and non-pole contributions in that theory also poses a serious problem
for the generalized GDH integrals.  As long as this problem exists,
the given comparison between different predictions should not be taken
too seriously.  However the large spread of the obtained values
certainly poses the challenge to solve the theoretical problem and to
measure the generalized GDH integrals a soon as possible. Since we
expect a negative slope of the purely transverse GDH integrals with a
minimum as low as $Q^2\simeq m_{\pi}^2$, it is quite fortunate that
some of the JLab experiments take data down to very low momentum
transfer.

We conclude that the presently running and newly proposed
experiments with beam and target/recoil polarization will give
invaluable information on the spin structure of the nucleon. It is
to be hoped that this quantitative increase of our knowledge will
also sharpen our theoretical tools to test the applicability and
the predictions of QCD in the confinement region.

\vspace{1.5cm}
\begin{center}
{\bf ACKNOWLEDGEMENT}\\
\end{center}
The authors gratefully acknowledge the support of the Deutsche
Forschungsgemeinschaft (SFB 443) and of the Heisenberg-Landau
program. We would like to thank the members of the GDH
Collaboration for many fruitful discussions, in particular H.-J.
Arends, R. Lannoy, and A. Thomas.

\newpage
\mediumtext
\begin{table}[htbp]
\begin{center}
\caption{Contributions to the GDH sum rule for the proton in units
of $\mu$b. Note that $\nu = 1.66 \, GeV$ corresponds to $W=2\, GeV
. $ } \vspace{0.5cm} \label{tab1}
\begin{tabular}{lccc}
energy range & channels & contributions & reference \\ \hline
$\nu_0<\nu<200$ MeV & $\pi^+n$ & 31 & MAID~\cite{Dre99a} \\ &
$\pi^0p$ & -1 &
\\ \hline 200 MeV$<\nu<800$ MeV & all &  -218$\pm$6 & MAMI~\cite{Ahr00}\\
\hline 800 MeV$<\nu<1.66$ GeV & $\pi^+n$ & -30 &
MAID~\cite{Dre99a}
\\ & $ \pi^0p$ & -6 & \\ \cline{2-4} & $\eta$ p & 7 & our estimate
\\ & $\pi \pi$ N & -15$\pm$ 8 & \\ & $\Lambda K^+, \Sigma K^+$ & 4
&
\\ \hline $\nu>1.66$ GeV & all & 26 $\pm$7 & Bianchi et al.~\cite{Bia99} \\
\hline total & all & -202 $\pm$ 10 &  \\ \hline \hline sum rule &
all & -204 & GDH~\cite{Ger65}
\end{tabular}
\end{center}
\end{table}

\begin{table}[htbp]
\begin{center}
\caption{Contributions to $\gamma_0$ and $\delta_0$ for the proton
in units of $10^{-4}$fm$^4$. The quantities with an asterisk are
evaluated by use of Eqs.~(\ref{eq30}) and~(\ref{eq30a}). For
details see text. } \vspace{0.5cm} \label{tab2}
\begin{tabular}{llddc}
contributions & & $\gamma_0$ & $\delta_0$ & reference \\ \hline
one-pion & l=0 &2.65  & 2.03 & MAID~\cite{Dre99a} \\
         & l=1 &-2.90  & -0.36      \\
         & l$\ge$ 2 &-0.39 &  -0.01  \\
others &$n\pi + \eta$ &-0.01 & 0.00& our estimate \\ \hline total
& & -0.65 & 1.66 &  \\ \hline \hline $\nu_0<\nu<200$ MeV & &1.04
&0.81 & MAID~\cite{Dre99a}
\\ 200 MeV $<\nu<800$ MeV & &-1.66 &0.82 &  \\ 800 MeV $<\nu<1.6$ GeV &
&-0.03 &0.02 & our estimate \\ \hline total & &-0.65 &1.66 & \\
\hline \hline ChPT $O(p^3)$ &  & 4.6 & 2.3$^{\ast}$ &
BKM~\cite{Ber95}
\\ \hline ChPT $O(p^4)$ & & -3.9 & 1.5$^{\ast}$ &
 JKO~\cite{JiO99}, KMB~\cite{Kum99}
 \\ & & -0.9 & 4.5$^{\ast}$ &
GHM~\cite{Gel00}
\end{tabular}
\end{center}
\end{table}

\begin{table}[htbp]
\begin{center}
\caption{Slopes of GDH related integrals for the proton at $Q^2=0$
in units of GeV$^{-2}$. The quantities with an asterisk are
evaluated by use of Eq.~(\ref{eq17}) and with $\gamma_0$ as given
by the respective authors. For details see text. }
 \vspace{0.5cm}
  \label{tab3}
\begin{tabular}{ddddddl}
    $I_1'$ & $I_2'$ & $I_3'$ & $I_A'$ & $I_B'$ &
$I_C'$ & reference\\
\hline  4.4 & -8.8 & -4.4 & -4.8 & -3.4 &
-5.6 & MAID~\cite{Dre99a} \\
   & -8.2 & & & & & BC~\cite{Bur70} \\ 0 &
& & 9.1 & 0$^{\ast}$ & 9.1 & $O(p^3)$ BKM~\cite{Ber93} \\
   7.0 & & & -14.4 & -6.6$^{\ast}$ &  & $O(p^4)$ JKO~\cite{JiK99}\\
3.4 & & & & & &  AIL~\cite{Ans89} \\
  8.0 & -8.2 & -0.2 & & & & ST~\cite{Sof93}
\end{tabular}
\end{center}
\end{table}

\vspace{1.5cm}

\end{document}